\documentclass[sigconf]{acmart}

\usepackage[T1]{fontenc}
\usepackage[utf8]{inputenc}
\usepackage{cleveref}
\usepackage{csquotes}
\usepackage{fancyvrb}

\usepackage{paralist}

\usepackage{booktabs}

\usepackage{textcomp}

\usepackage{lipsum}
\usepackage{balance}
\usepackage{xcolor}
\usepackage{xspace}

\usepackage{uri}
\usepackage{layouts}

\newcommand{\afblock}[1]{\noindent{\textbf{#1 }}}
\newcommand{\takeaway}[1]{\noindent{\textbf{Takeaway.}} \textit{#1}}

\newenvironment{smallenumerate}{
\begin{enumerate}
\setlength{\parsep}{0.3ex}
\setlength{\leftmargin}{0.1pt}
\setlength{\listparindent}{-8ex}
\setlength{\itemsep}{0.2ex}
\setlength{\topsep}{0ex}
\setlength{\parskip}{0ex}
\setlength{\partopsep}{0ex}
}{
\end{enumerate}
}

\usepackage{enumitem}
\usepackage{xcolor}

\usepackage{xspace}
\usepackage{acro}
\usepackage{mfirstuc}
\acsetup{list-long-format=\capitalisewords, list-style=longtable}
\DeclareAcronym{cdn}{
	short = CDN,
	long = content delivery network
}
\DeclareAcronym{as}{
	short = AS,
	long = autonomous system
}
\DeclareAcronym{isp}{
	short = ISP,
	long = Internet service provider
}

\newcommand{\eg}{e.g., }

\newcommand{\ie}{i.e., }

\newcommand{\applecdn}{Apple Meta-CDN}

\newcommand{\offload}{\textit{offload}\xspace}
\newcommand{\overflow}{\textit{overflow}\xspace}
\newcommand{\Offload}{\textit{Offload}\xspace}
\newcommand{\Overflow}{\textit{Overflow}\xspace}

\newcommand{\srcAS}{\textit{Source AS}}
\newcommand{\handoverAS}{\textit{Handover AS}}

\copyrightyear{2018} 
\acmYear{2018} 
\setcopyright{acmlicensed}
\acmConference[IMC '18]{2018 Internet Measurement Conference}{October 31-November 2, 2018}{Boston, MA, USA}
\acmBooktitle{2018 Internet Measurement Conference (IMC '18), October 31-November 2, 2018, Boston, MA, USA}
\acmPrice{15.00}
\acmDOI{10.1145/3278532.3278567}
\acmISBN{978-1-4503-5619-0/18/10}

\begin{document}

\title{Dissecting Apple's Meta-CDN during an iOS Update} 

\author{Jeremias Blendin}
\orcid{0000-0002-1773-4209}
\affiliation{%
  \institution{Technische Universität Darmstadt / DE-CIX}
}

\author{Fabrice Bendfeldt}
\affiliation{%
  \institution{Technische Universität Darmstadt}
}

\author{Ingmar Poese}
\authornote{Corresponding authors:\\
Jeremias~Blendin: jblendin@kom.tu-darmstadt.de, 
Ingmar~Poese: ipoese@benocs.com,
Oliver~Hohlfeld: oliver@comsys.rwth-aachen.de}
\affiliation{%
  \institution{BENOCS}
}

\author{Boris Koldehofe}
\affiliation{%
  \institution{Technische Universität Darmstadt}
}

\author{Oliver Hohlfeld}
\orcid{1234-5678-9012}
\authornotemark[1]
\affiliation{%
  \institution{RWTH Aachen University}
}

\renewcommand{\shortauthors}{J. Blendin et al.}

\begin{abstract}
Content delivery networks (CDN) contribute more than 50\% of today's Internet traffic.
Meta-CDNs, an evolution of centrally controlled CDNs, promise increased flexibility by multihoming content.
So far, efforts to understand the characteristics of Meta-CDNs focus mainly on third-party Meta-CDN services.
A common, but unexplored, use case for Meta-CDNs is to use the CDNs mapping infrastructure to form self-operated Meta-CDNs integrating third-party CDNs.
These CDNs assist in the build-up phase of a CDN's infrastructure or mitigate capacity shortages by offloading traffic.
This paper investigates the Apple CDN as a prominent example of self-operated Meta-CDNs.
We describe the involved CDNs, the request-mapping mechanism, and show the cache locations of the Apple CDN using measurements of more than 800 RIPE Atlas probes worldwide.
We further measure its load-sharing behavior by observing a major iOS update in Sep. 2017, a significant event potentially reaching up to an estimated 1 billion iOS devices.
Furthermore, by analyzing data from a European Eyeball ISP, we quantify third-party traffic offloading effects and find third-party CDNs increase their traffic by 438\% while saturating seemingly unrelated links. 
\end{abstract}

\maketitle

\section{Introduction}
\label{sec:intro}

\Acp{cdn} have become a key component of the Internet~\cite{adhikari2012tale,calder2015anycast}.
In order to reduce latencies and increase the availability of content for its consumers  they serve content from nearby servers, which has flattened the hierarchical structure of the Internet~\cite{labovitz2010internet}.
\acp{cdn} achieve this by providing three major functions: geographically distributed content caches, direct connections to ISPs or IXPs, and a request mapping to select the best cache location.

Unsurprisingly, \acp{cdn} cause high traffic shares: \eg more than half of the traffic of a North American~\cite{gerber2011backbone} and a European~\cite{poese2010padis} \ac{isp} can be attributed to only a few \acp{cdn}.
Despite these characteristics, customers of a single \ac{cdn} are bound to its cost model, performance, and geographic distribution---a limitation solved by publishing content on {\em multiple} \acp{cdn}, which requires an additional request mapping for \ac{cdn} selection.

We refer to \ac{cdn} selecting infrastructures as Meta-CDN services.
Known instances of third-party Meta-CDNs~\cite{frank2013pushing, xue2017cdn, mukerjee2016broker, BruceBroker17} provide services to implement custom and dynamic request mapping policies to direct traffic to the different \acp{cdn}. 
The characteristics of third-party Meta-CDNs have been studied in part on the example of Conviva~\cite{conviva,dobrian2011conviva,mukerjee2016broker} and Cedexis~\cite{cedexis-pam} as the prevalent operators.

As an alternative to using third-party Meta-CDN providers, large content providers often prefer to build up their own infrastructure and only depend on third-party \acp{cdn} when necessary. 
This approach leads to a hybrid model where a content provider uses third-party \acp{cdn} to supplement its own infrastructure, \eg to handle overload.
In this model, the content provider effectively becomes a {\em self-operated Meta-CDN} by directing traffic either to its own infrastructure or to third-party \acp{cdn}.
However, little is known about this type of Meta-CDN and its implications.

In this paper, we shed light on self-operated Meta-CDNs using a detailed analysis of the \applecdn{} as a prominent example.
It is used to deliver Apple services (\eg iTunes) and to deliver Apple software updates (\eg iOS updates for up to 1\,billion devices including the iPhone, iPad, and iPod~\cite{AppleAnnual16}).
The rollout of major updates only happens a few times a year but for all devices at the same time, which creates substantial traffic demands.
In this study, we show how the \applecdn{} handles these update events by examining a major iOS update.
From this, we discover that the \applecdn{} indeed operates a Meta-CDN service by involving third-party \acp{cdn} in addition to their own \ac{cdn} infrastructure.
We further demonstrate the {\em consequences} of this update event and particularly the Meta-CDN service {\em on ISP traffic} by analyzing detailed traffic traces from a European Eyeball ISP.
To reason about the effects of the Meta-CDN service, we correlate the RIPE measurement with the ISP dataset.
Our contributions are as follows:
\begin{smallenumerate}
	\item We provide the first characterization of a self-operated Meta-CDN.
We describe the involved CDNs, the request-mapping mechanism, and discover the cache locations of the Apple CDN using more than 800 RIPE Atlas~\cite{RipeAtlas} probes worldwide.
	\item We observe its request mapping and load sharing behavior during a major iOS update in Sept. 2017.
	\item We provide the first study on the impact of a Meta-CDN service on ISP traffic.
 In this study, we cross-correlate RIPE Atlas DNS measurements with ISP traffic data to quantify the effect of offload and overflow traffic for the ISPs with regards to the Apple iOS update.
We find third-party CDN traffic spikes to reach 438\%, that the distribution of offload traffic is dynamic on a daily basis and that, due to overflow, seemingly unrelated links suddenly saturate.
\end{smallenumerate}

\section{Background \& Related Work}
Building on a geographically distributed infrastructure, \acp{cdn}~\cite{adhikari2012tale,calder2015anycast} enable high availability and low latency when content is retrieved from close-by servers.
Multiple studies contribute towards an understanding of these infrastructures, including their performance~\cite{adhikari2012tale,calder2015anycast,nygren2010akamai,otto2012cdn}, and their request mapping mechanisms~\cite{su2009drafting,chen2015mapping}.
Further optimizations of request mapping mechanisms are proposed, \eg based on anycast~\cite{chen2015mapping,flavel2015fastroute}, by enabling ISP-CDN collaboration~\cite{poese2010padis}, or both~\cite{wichtlhuber17}.
Using multiple \acp{cdn} can offer cost reductions (\eg due to  different prices to serve content at different times, or due to traffic volume contracts) and higher availability (\eg in high-load scenarios). 
This case of \emph{multi-homed} content---\ie content that is served by multiple \acp{cdn}---requires an additional request mapping layer to route requests to \acp{cdn} selected for delivery~\cite{frank2013pushing,liu2012optimizing,xue2017cdn}.

A Meta-CDN service can be provided by third-party infrastructures, most prominently Conviva for video delivery~\cite{conviva, dobrian2011conviva} and Cedexis for general (web) services~\cite{cedexis, cedexis-pam, xue2017cdn}. 
Within these Meta-CDNs, the request mapping function of a participating content provider is unknown.
Hence, the demand and corresponding traffic are harder to predict for Eyeball ISPs.
This lack of information can cause unexpected traffic skews as well as significant variance in demand over short timescales~\cite{mukerjee2016broker}.
Understanding the request mapping potential and its consequences on traffic---as addressed by our study---is thereby needed to understand Meta-CDNs and their consequences on traffic better.

\section{Dissecting the Apple CDN}
\label{sec:applecdn}
We start our analysis by dissecting the operating principle of the \applecdn{} and will show {\em i)} how iOS devices discover and download updates, {\em ii)} how download requests are mapped to CDNs, and {\em iii)} the architecture of Apple's content cache infrastructure.

\subsection{iOS Device Behavior}
\label{sec:device}

We examine the iOS update discovery and download behavior by analyzing traffic from an Apple TV and an iPhone 7 Plus device.
We found that iOS devices download two manifest files from \emph{mesu.apple.com} once per hour to check for available updates.
The first file ~\cite{apple_softwareupdateurl}, termed manifest, contains the version and download URL for every device and OS version combination with about 1800 entries as of July 2017, and the second file~\cite{apple_updatebrainurl} contains only six entries.
We did not observe the second file being used in the collected data, and therefore, assume it to be a last-resort mechanism that enables devices with outdated iOS software to upgrade their software.
If the manifest contains information on a new software update, the user is notified on its availability.
When the user manually initiates the update process, the update file is downloaded from \emph{appldnld.apple.com} via HTTP.

\subsection{CDN Selection Process}
\label{sec:request-mapping}
The Apple iOS update CDN is a Meta-CDN that relies on a mixture of its own infrastructure and infrastructure from other CDN providers for both the request mapping and the file delivery.
The request mapping in Apple's self-operated Meta-CDN uses location-based dynamic DNS resolution~\cite{su2009drafting,VivisectingYoutube}.

\begin{figure}[ht]
  \centering
  \includegraphics[width=0.85\columnwidth]{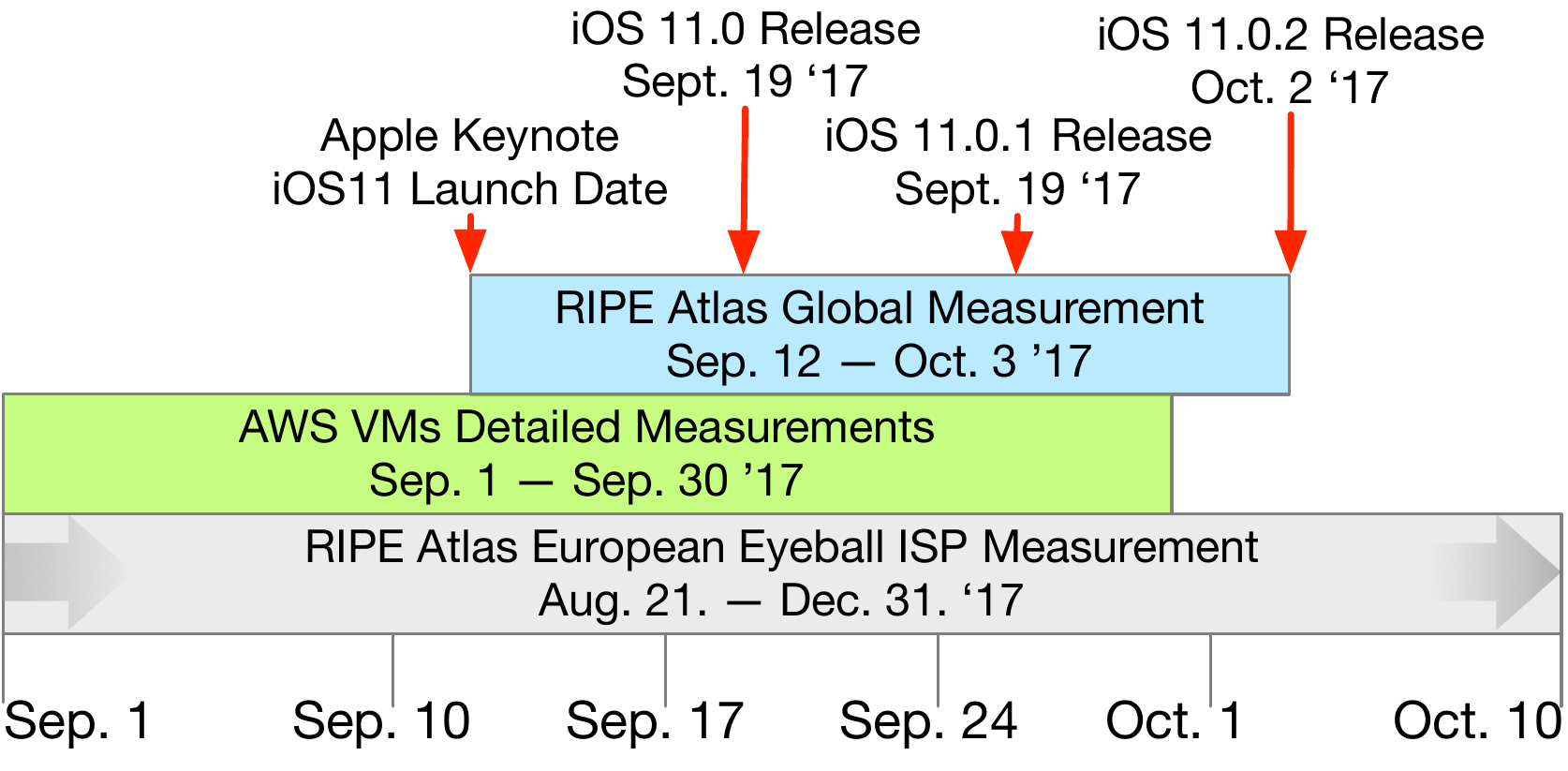}
  \caption{Active measurement timeline.}
  \label{fig:timeline}
\end{figure}

\afblock{Measurement Setup.}
To account for the location-dependence in the request mapping, we queried the \applecdn{} from globally distributed RIPE Atlas~\cite{RipeAtlas} probes and Amazon AWS VMs.
We show the measurement period in Figure~\ref{fig:timeline}.
Full recursive DNS resolution measurements and checking the availability of the relevant files on the Apple CDN servers was done on nine AWS VMs on every continent except Africa.
To understand the global CDN request mapping behavior in detail, we monitored the \applecdn{} from 800 globally distributed RIPE Atlas probes issuing DNS requests and traceroute probes.
In contrast to the AWS VMs, these probes only collect the DNS reply data.
These Atlas probes issued DNS requests every 5 minutes during one week before and after the iOS 11 update released on Sep. 19, 2017; the data is publicly available \cite{RipeAtlasios11}.
We perform traceroutes to all server IPs identified via DNS every hour.
Finally, to understand the specific behavior of the \applecdn{} from the viewpoint of the European Eyeball ISP studied in Section~\ref{sec:overflow}, an additional 400 Atlas probes are used, dedicated to measuring inside the ISP every 12 hours between Aug. 20 and Dec. 31, 2017.
The measurement approach is designed to capture the diversity of DNS request mappings using the RIPE Atlas probes and the DNS mapping infrastructure as well as the availability of the content using AWS VMs.
The approach is generic, which means it could be applied to any other CDN.

\begin{figure}[tb]
  \centering
  \includegraphics[width=0.9\columnwidth]{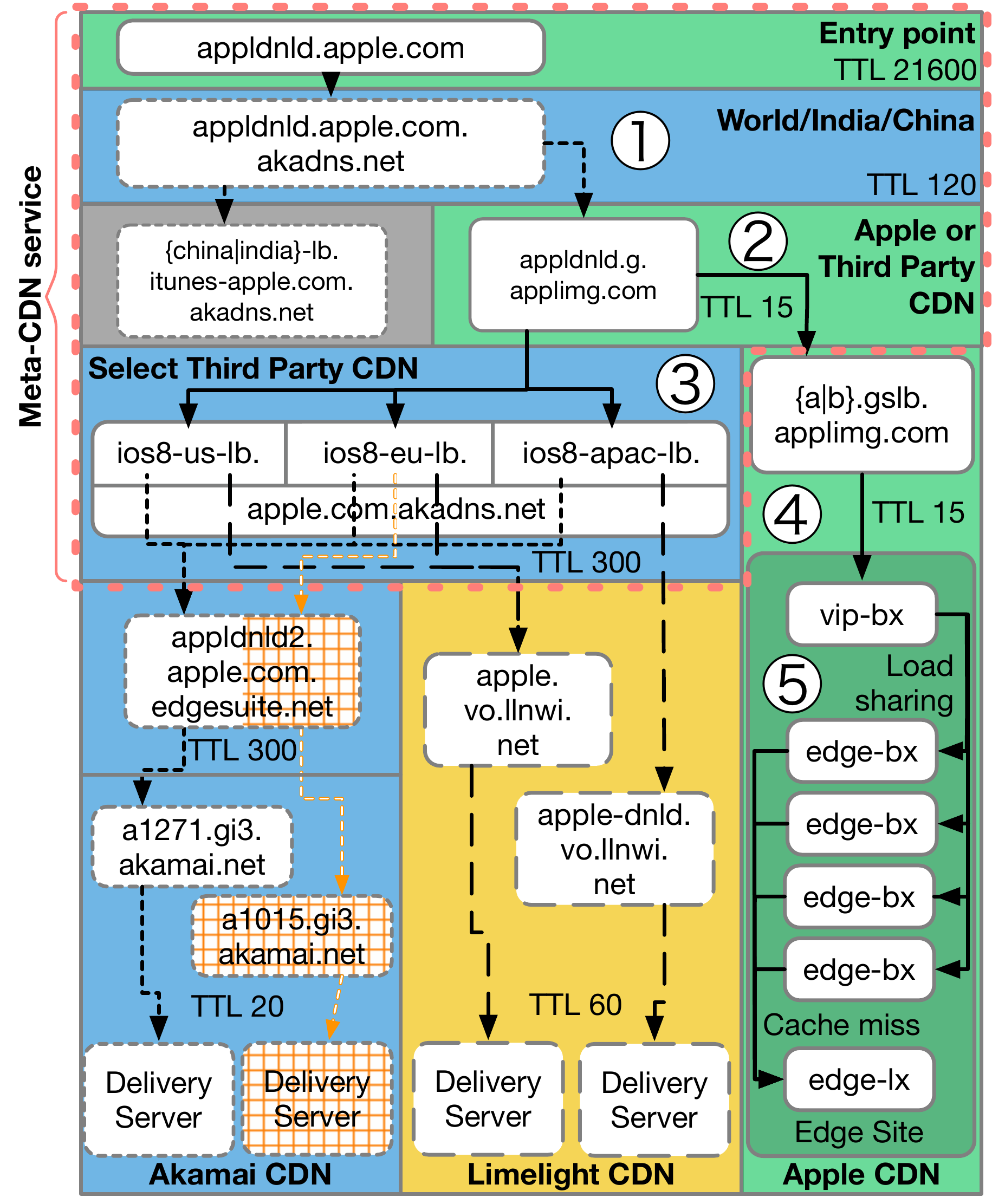}
  \caption{Request mapping DNS and load sharing infrastructure.}
  \label{fig:dns-infrastructure-overview}
\end{figure}

\afblock{Apple CDN Request Mapping Infrastructure.}
In Section~\ref{sec:device}, we observed \emph{appldnld.apple.com} to be the entry point for downloading update images on iOS devices.
By using our measurements, we dissect the involved request mapping infrastructure, as shown in Figure~\ref{fig:dns-infrastructure-overview}.
The parts depicted in an orange checker pattern are modifications observed during the rollout of the iOS 11.0 (see \Cref{sec:ios11-update-rollout}).
The depiction contains the parts of the request mapping process where decisions are made.
Each box, except for \textcircled{5}, represents a DNS name.
Each arrow represents a \texttt{CNAME} redirect with the time-to-live (TTL) in seconds of the following DNS name--which we found to be stable throughout our measurements. 
The part named edge site \textcircled{5} illustrates the cache server structure of Apple's own CDN and is discussed in \Cref{sec:apple-deliver-servers}.

The first step \textcircled{1} uses an Akamai DNS service to differentiate whether the request originates from India or China, or from other countries.
If the request does not originate from India or China, it is sent back to the Apple infrastructure (step \textcircled{2}).
We speculate that this decision is driven by infrastructure availability in these regions.
Since the density of RIPE probes in these regions is low, we do not study these regions further.

The CDN selection (Meta-CDN service)---\ie whether to serve the request by Apple's own CDN or by a third-party CDN---is performed in step \textcircled{2}.
In this step, the selection of the \ac{cdn} is provided by the DNS resolution of \emph{appldnld.g.applimg.com}.
This DNS \texttt{CNAME} has a TTL of 15\,s to enable quick reroutes.
If the Apple \ac{cdn} is selected \textcircled{4}, a final redirection is performed by the two \emph{\{a|b\}.gslb.applimg.com} DNS entries, which results in IPs of Apple CDN cache nodes.
The name \emph{gslb} suggests that it functions as a global server load balancer.
We dedicate \Cref{sec:apple-deliver-servers} to discuss Apple's own \ac{cdn} infrastructure in detail.

If a third-party CDN is selected for delivery \textcircled{3}, the request is forwarded back to Akamai's DNS infrastructure.
We found that Akamai provides three load-balance DNS entries for conducting the selection of the third-party CDN: \emph{ios8-\{eu|us|apac\}-lb.apple.com.akadns.net}.
Depending on the region, different third-party CDNs were used:
\begin{inparaenum}[\em i)]
  \item US: Akamai, Limelight, Level3
  \item EU: Akamai, Limelight, Level3
  \item APAC: Akamai, Limelight.
\end{inparaenum}
The DNS handover points to the Akamai and Level3 CDNs are the same for US, EU, and APAC.
However, Limelight uses one DNS specifically for the the US (\emph{apple.vo.llnwi.net}) and one for APAC (\emph{apple-dnld.vo.llnwd.net}).
Level3 was removed from the request mapping in late June 2017 and is, therefore, not included in \Cref{fig:dns-infrastructure-overview}.

Our interpretation of the request mapping design is twofold: first, it provides easy control over the distribution shares of third-party CDNs where alternatives are available, and second, the coverage of areas where Apple has not deployed its own infrastructure. 
The modification of distribution shares of third-party CDNs was observed in our measurements.
The control of the distribution shares are directly controlled by Apple and we assume are driven by commercial interests. 
Akamai's infrastructure is used at two crucial points in the Meta-CDN service: the decision between China and India and the rest of the world \textcircled{1}, and the selection of third-party CDNs \textcircled{3}.
In both cases, the decision may result in directing the request to a third-party CDN.
We speculate that Akamai is used because, in contrast to Apple's, its infrastructure is available globally.
In general, Akamai seems to be used when the lack of local mapping infrastructure in some regions may impact the performance of the DNS resolution process. 
Apple's involvement in \textcircled{1} and \textcircled{2} is required, although this might have an impact on the mapping performance, to ensure that Akamai as a mapping provider can be replaced if necessary.
In conclusion, we find the design's primary goal is to ensure Apple's bargaining power with its CDN suppliers.

Finally, we found that none of the mapping entry points responds to requests for IPv6 resolution; only IPv4 is used.

\takeaway{The delivery of Apple updates relies on a Meta-CDN to select cache delivery infrastructures of which Apple operates one on its own. Most notably, the Meta-CDN involves three selection (mapping) steps of which two are run by Akamai and one by Apple.}

\begin{figure}[t]
  \centering
  \includegraphics[width=1.0\columnwidth]{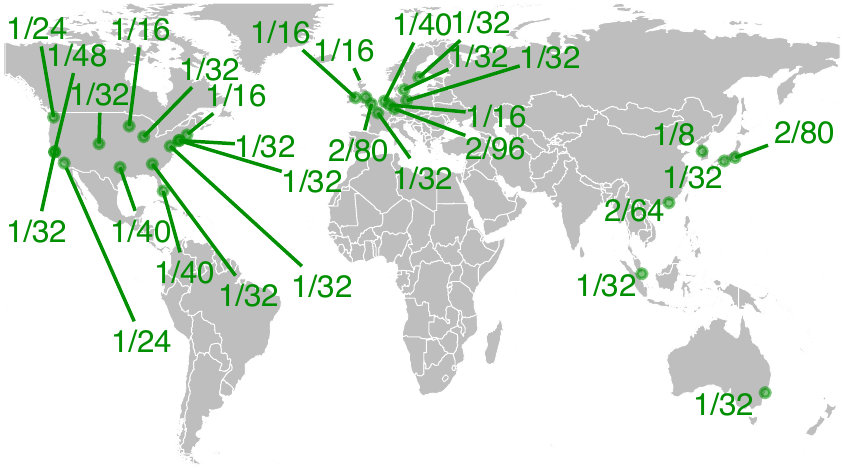}
  \caption{Apple delivery server locations.}
  \label{fig:serverlocations}
\end{figure}

\subsection{Apple's Own CDN Infrastructure}
\label{sec:apple-deliver-servers}

We now discuss Apple's \ac{cdn} infrastructure in detail (see \textcircled{4} in Figure~\ref{fig:dns-infrastructure-overview}).
From the performed DNS resolutions, we identified Apple's delivery servers to use the \emph{17.253.0.0/16} subnet and to have reverse DNS names such as \emph{usnyc3-vip-bx-008.aaplimg.com}.
By scanning Apple's IP range (17.0.0.0/8) for the availability of iOS image downloads and by enumerating the DNS names using the \emph{Aquantone} tool \cite{aquatone}, we reconstructed the Apple's naming scheme, which involves the identifiers shown in \Cref{tab:apple-cdn-server-names}.
The location naming is consistent with the UN/LOCODE scheme~\cite{locode}, except for one location, London, \emph{uklon} which should be \emph{gblon}.
By using the naming scheme information, we identified the locations and functionality of Apple CDN cache sites. 
\Cref{fig:serverlocations} shows the 34 discovered Apple CDN delivery site locations, known as \emph{edge} sites in the server naming scheme.
Their IPs are distributed via request mapping system to clients, in the depiction their labels denote \emph{<\# of sites>/<total \# of cache servers>}.
One desired property of a \ac{cdn} is to have a globally distributed set of edge servers to serve nearby clients.
\begin{table}[h]
\footnotesize
\centering
\caption{Apple server naming scheme}
Naming Scheme: \texttt{ab-c-d-e}.aaplimg.com\\
Example: \emph{usnyc3-vip-bx-008.aaplimg.com}
\begin{tabular}{lp{6.5cm}}
  \toprule
  Identifier & Meaning \\
  \hline
  a & UN/LOCODE location (\eg deber for Berlin) \\
  b & Location site id (\eg 1)\\
  c & Function: vip, edge, gslb, dns, ntp and tool  \\
  d & A secondary function identifier: bx, lx and sx  \\
  e & Id for same function server (\eg 004) \\
  \bottomrule
\end{tabular}
\label{tab:apple-cdn-server-names}
\end{table}

Interestingly, the {\em internal structure} of {\em edge sites} can be revealed by analyzing \emph{HTTP} header information used during downloads; an example of the relevant part of the header is shown below:
\begin{Verbatim}[fontsize=\footnotesize]
X-Cache: miss, hit-fresh, Hit from cloudfront
Via: 1.1 2db316290386960b489a2a16c0a63643.cloudfront.net (CloudFront),
 http/1.1 defra1-edge-lx-011.ts.apple.com (ApacheTrafficServer/7.0.0),
 http/1.1 defra1-edge-bx-033.ts.apple.com (ApacheTrafficServer/7.0.0)
\end{Verbatim}
From that, we infer that client requests are directed to nodes with the \emph{vip-bx} function that forwards requests to one of four associated nodes with the \emph{edge-bx} function as denoted by \textcircled{5} in \Cref{fig:dns-infrastructure-overview}.
If the file is not found there, the request is forwarded to a node with the \emph{edge-lx} function.
The term \emph{vip} suggests \enquote{virtual IP} and that the associated server is a load balancer for four \emph{edge-bx} nodes.
The use of load balancers in the delivery sites instead of purely relying on DNS suggests that a single Apple CDN IP represents the download capacity of four servers.
This is why the number of servers per location in \Cref{fig:serverlocations} refers to the number of \emph{edge-bx} nodes.

\takeaway{Besides using third-party \acp{cdn}, Apple has created a substantial cache infrastructure of CDN sites.
Their density of sites is the highest in the USA followed by Europe and East Asia, while the South American and African continents lack distribution data centers.}

\section{Characterizing the Apple CDN \\ during an iOS Update Rollout}
\label{sec:ios11-update-rollout}

To study the behavior of the \applecdn{} during operation {\em and} under load, we measured the rollout of iOS 11, a major iOS software update released by Apple on Sep. 19, 2017 at 17h UTC.
For this perspective, we use the RIPE Atlas DNS measurements described in Section~\ref{sec:request-mapping}, which started on Sep. 12---7 days before the update was made available to iOS users (see Figure~\ref{fig:timeline}).
\begin{figure}[t]
    \includegraphics[width=1.0\columnwidth]{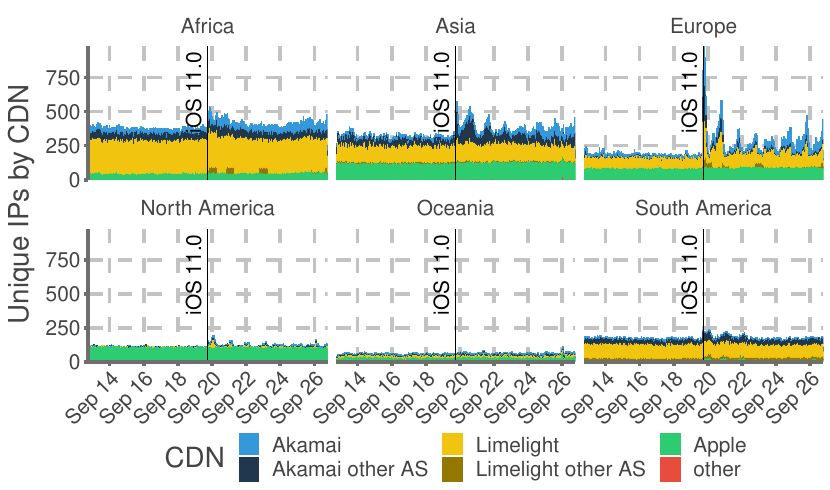}
    \caption{Number of unique CDN cache IPs of the worldwide measurement.}
    \label{fig:updatecacheips-continents}
\end{figure}

To provide a global perspective of the changes in the CDN infrastructure during the update, we show the involved \acp{cdn} and their number of unique server IPs per continent in Figure~\ref{fig:updatecacheips-continents} as seen in responses to our DNS queries.
Cache IPs that are used by Akamai or Limelight but not located within their respective \acp{as} are denoted as \enquote{other AS}.
The focus of this discussion is the changes in the Meta-CDN service, specifically the choices of the Apple Meta-CDN operators when it comes to selecting third-party \acp{cdn}.
As discussed in \Cref{sec:request-mapping}, we assume commercial interests to be the driving factor for Apple's Meta-CDN service design.
Information on the actual traffic caused by the Apple Meta-CDN during the event is not available globally but for a single European Eyeball ISP.
The results of which are discussed in \Cref{sec:overflow}.

\afblock{Global perspective.}
As expected from the distribution of Apple CDN sites, North America features the highest ratio of Apple cache IPs, while South America and Africa show the highest ratio of third-party CDN IPs.
Interestingly, despite the high number of Apple CDN sites in Europe, $\sim50\%$ of  the observed IPs belong to third-party \acp{cdn}.
Furthermore, Europe is the only continent that shows a considerable spike in the number of unique IPs following the update event.
We observe a maximum of 977 IPs immediately after the iOS software release on Sep. 19 at 18h UTC.
This peak is more than four times the average of 191 of unique cache IPs that were observed in the two days before.
The major part of the increase in unique IPs is caused by Limelight and, to a lesser part, Akamai, with the latter increasing the number of cache IPs that are located in third-party networks.
Akamai's increased load correlates with the appearance of a new \texttt{CNAME}: \ie six hours after the update started on Set. 19 around 23h, \emph{a1015.gi3.akamai.net} was added in the Akamai CDN for requests coming from \emph{ios8-eu-lb.apple.com.akadns.net}.
We assume the reason for this behavior, in comparison to North America, is twofold.
First, Apple's infrastructure is not as built out yet in Europe. 
Second, according to estimates by the APNIC~\cite{apnicixp}, the ISP market in the USA is more consolidated with roughly 60\% market share for the ten largest ISPs, while in Europe the ten largest ISPs only have a market share of about 30\%.
The higher market fragmentation in Europe requires a more complex CDN deployment.

The new delivery path is depicted in an orange checker pattern in \Cref{fig:dns-infrastructure-overview}.
This observation shows that the \applecdn{} handles the increased traffic demand of the update by delegating traffic to third-party \acp{cdn} using their Meta-CDN service.
We did not observe any proactive changes to Apple's request mapping infrastructure before the release of the update.

\begin{figure}[t]
    \includegraphics[width=1.0\columnwidth]{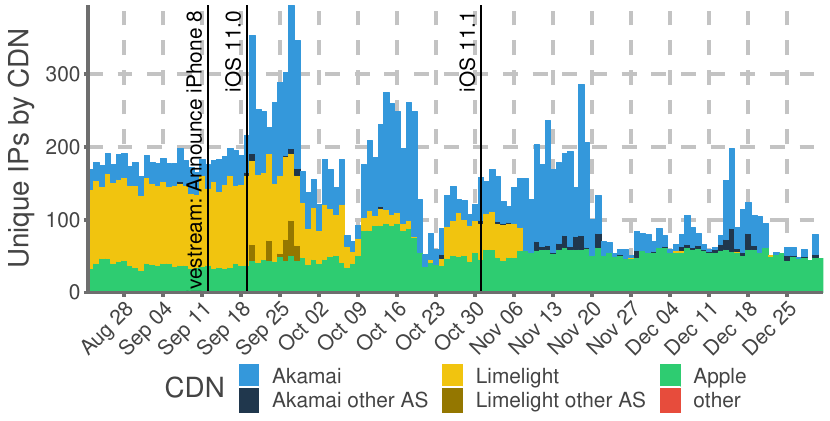}
    \caption{Number of unique CDN cache IPs of the European Eyeball ISP measurement.}
    \label{fig:ripe-result-dtdauer-unique}
\end{figure}
\afblock{European ISP.}
We show the same measurement performed by 400 RIPE Atlas probes located in the network of the European Eyeball ISP studied in Section~\ref{sec:overflow} in Figure~\ref{fig:ripe-result-dtdauer-unique}.
Here, the load increase by the update is not distributed equally among all \acp{cdn}, and the spike in the number of IPs is not as pronounced as in the whole of Europe.
Still, we observed an increasing number of IPs for other \acp{cdn}, most notably the number of Akamai CDN IPs rise by 408\% from Sep. 18 to Sep. 20.
We find Apple's CDN to have a somewhat stable number of IPs that are not increased, \eg suggesting that Apple's CDN cannot further increase the number of download cache locations.
The steady number of Apple IPs also indicates that the demand could not be satisfied with their infrastructure alone and thus needed to be delegated to third-party \acp{cdn}---most notably Akamai.
This assumption is supported by the fact that it e.g., takes six hours for Akamai to increase its number of distributed IP addresses to its load-dependent peak.
The ISP is thus facing the challenge of the update to cause traffic spikes by third-party \acp{cdn}---solely because of Apple's Meta-CDN service.
This motivates us to study consequences for the ISP in the next section.

\takeaway{The rollout of a major software update causes high traffic demands that are handled by offloading requests to third-party \acp{cdn} using the Meta-CDN service. During the update, no party has full control over the entire infrastructure while user assignment and server selection is made by multiple \acp{cdn} independently.}

\section{ISP Perspective on a Meta-CDN}
\label{sec:overflow}

The previous section showed that the \applecdn{} handles peak traffic events by involving third-party \acp{cdn}.
This process, known as offloading, is controlled by the Apple operated Meta-CDN service and has effects on ISP traffic engineering and their peering links.
These effects motivate us to look at a Meta-CDN service from the perspective of a Tier-1 European Eyeball ISP.

\begin{figure}[tb]
  \centering
  \includegraphics[width=0.8\columnwidth]{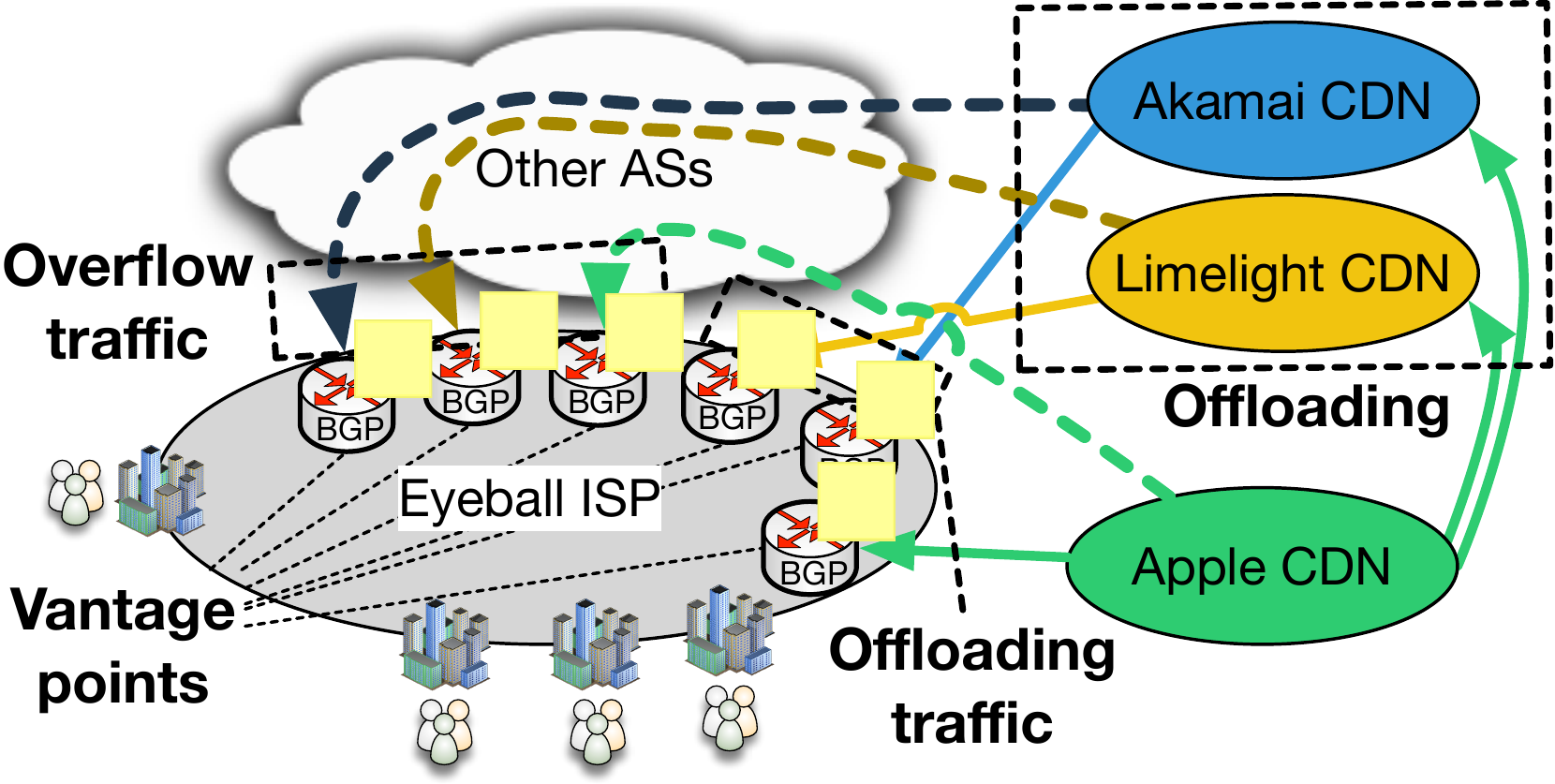}
  \caption{Illustration of CDN offloading and traffic overflow from perspective of an ISP.}
  \label{fig:offloading-overflow}
\end{figure}

\subsection{Definition of \offload{} and \overflow{} Traffic}
For ease of presentation, we define the traffic sources as follows.
\begin{itemize}[topsep=0pt,parsep=0pt,partopsep=0pt,noitemsep]
  \item \textbf{\srcAS} is the \ac{as} that originates the traffic of a connection, i.e., the \ac{as} of the servers' IP address.
	For example, the \srcAS{} for an Apple update can be Akamai if the request was delegated there by the \applecdn{} service.

  \item \textbf{\handoverAS} is the direct neighbor \ac{as} handing traffic to the measured ISP network.
	This AS can be completely unrelated to any used \ac{cdn} (\eg a transit \ac{as}).
\end{itemize}

The terms offload and overflow are defined as follows.
\begin{itemize}[topsep=0pt,parsep=0pt,partopsep=0pt,noitemsep]
  \item \textbf{\Offload} describes traffic that the \applecdn{} delivers via third-party \textit{CDN servers}, \ie the third-party \ac{cdn} is the \srcAS{}.
	In Figure~\ref{fig:offloading-overflow}, all traffic origination from Akamai and Limelight to the Eyeball ISP is offload traffic, as Apple hands that traffic over to third-party \acp{cdn}.
  \item \textbf{\Overflow} describes traffic received from non-direct neighbors, \ie the \srcAS{} and handover AS differ.
	In Figure~\ref{fig:offloading-overflow} this is all traffic received via ``\textit{Other ASes}''.
\end{itemize}          
Note that offload and overflow traffic are orthogonal.
For example, Akamai and Limelight traffic going via \textit{Other ASes} is both, offload and overflow traffic.
Apple traffic going via \textit{Other \textit{Other ASes}} is overflow traffic only.

\afblock{Consequences for ISPs.}
Both traffic types, offload and overflow, pose a significant strain on the \acp{isp} network.
For example, offload makes it harder for \acp{isp} to predict how much traffic to expect on what links.
This is because it is unclear {\em i)} which \acp{cdn} are selected by the Meta-CDN and {\em ii)} which \ac{cdn} is serving how much traffic.
Overflow further challenges traffic flow prediction, since seemingly unrelated peering links change their traffic volume.
And finally, offload is controlled by the Meta-CDN operator while overflow is handled by the individual \acp{cdn}'s load balancers. 
As these are two independent control loops, it becomes tough to predict how the 
traffic is going to behave.

\subsection{ISP Measurement Setup}

To quantify the effect of offloading and overflow, we gather BGP, Netflow and SNMP data directly on all border routers of a Tier-1 European Eyeball ISP (see vantage points in Figure~\ref{fig:offloading-overflow}) between Sep. 15 and Sep. 23. 
In total, we collect $\sim$300 billion Netflow records as well as $\sim$350 Million SNMP measurements while actively keeping track of $\sim$60 million  BGP routes in $\sim$300 active sessions.
We also get information about all active peering links and their respective \handoverAS.  
Furthermore, we verify that that internal cache links are handled as direct connections to the CDN controlling the cache. 

\subsection{Offload Impact}

{\em Offloading} happens when traffic from Apple is served via a third-party \ac{cdn}.
To quantify offloading, we select all \ac{cdn} server IPs observed in RIPE Atlas DNS measurements to the \applecdn{} located within the ISP (see Section~\ref{sec:ios11-update-rollout}) and cross-correlate them with Netflow for traffic flows and BGP for finding the Source AS.
Finally, we scale the Netflow traffic on the peering links by the byte Counters from SNMP to minimize Netflow sampling errors.
Thereby, we estimate the traffic caused by Apple and its offload \acp{cdn} due to the iOS update.

\begin{figure}[t]
  \includegraphics[width=1.0\columnwidth]{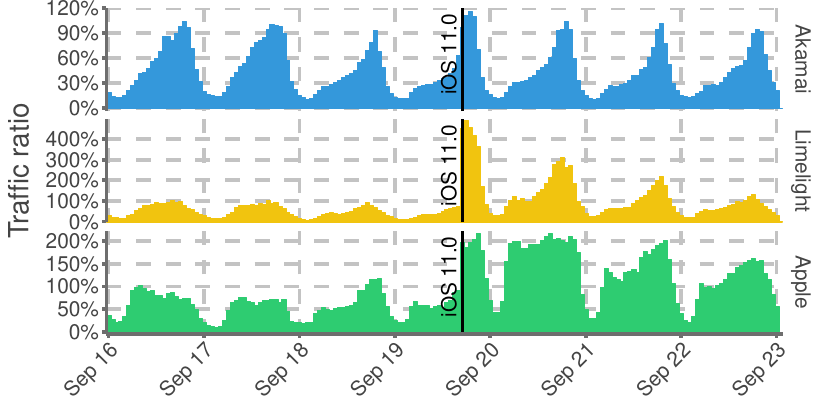}
  \caption{Update Traffic by source AS during the iOS update.}
  \label{fig:iOS-update-offload}
\end{figure}

Figure~\ref{fig:iOS-update-offload} shows the development of {\em offload} traffic as its {\em ratio} relative to the average peak traffic three days before the update for each CDN.
Here, a ratio of $100\%$ reflects the maximum traffic rate seen for a \ac{cdn} over the course of three days before the update.
A ratio of $> 100\%$ indicates traffic spikes, which we are interested in studying. 
We observe the maximum traffic from the Apple AS to peak at 211\%, from Limelight by 438\% and Akamai by 13\% on Sep. 19. 
When comparing the excess volume of extra traffic caused by the iOS update between the \acp{cdn}, we find that for, Sep. 19, 33\% come from Apple, 44\% from Limelight and 23\% from Akamai.
For Sep. 20 and 21 the bulk of the additional traffic is transported by Apple itself ($\sim$60\%) and Limelight ($\sim$40\%) with no additional Akamai traffic during this time. 

With Akamai being the biggest CDN traffic-wise, the assumption of Akamai carrying the bulk of the offload traffic could be assumed. 
However, this is not the case, as Akamai is only used on the Sep. 19, while all other days are shared between Apple and Limelight only. 
When looking at the traffic of Apple and Limelight, it also becomes visible that Apple runs at high capacity all of Sep. 20, while the other CDNs show a diurnal traffic pattern. 
This leads to the conclusion that Apple uses its own \ac{cdn} first before offloading. 

\takeaway{The observed traffic patterns show that knowledge about how CDNs possibly deliver the update says little about the distribution of how the traffic is split between them. Predictions become even harder when the traffic amounts are not stable and change on a daily basis.}

\label{sec:thirdpartyoverflow}

\subsection{Overflow Impact}

\setlength{\fboxsep}{0pt}
\setlength{\fboxrule}{0pt}

\definecolor{ASa}{HTML}{abdda4}
\newcommand{\asA}{AS \colorbox{ASa}{\makebox[1.4\width]{\strut\textcolor{black}{A}}}\xspace}
\definecolor{ASb}{HTML}{173B59}
\newcommand{\asB}{AS \colorbox{ASb}{\makebox[1.4\width]{\strut\textcolor{white}{D}}}\xspace}

\begin{figure}
  \includegraphics[width=1.0\columnwidth]{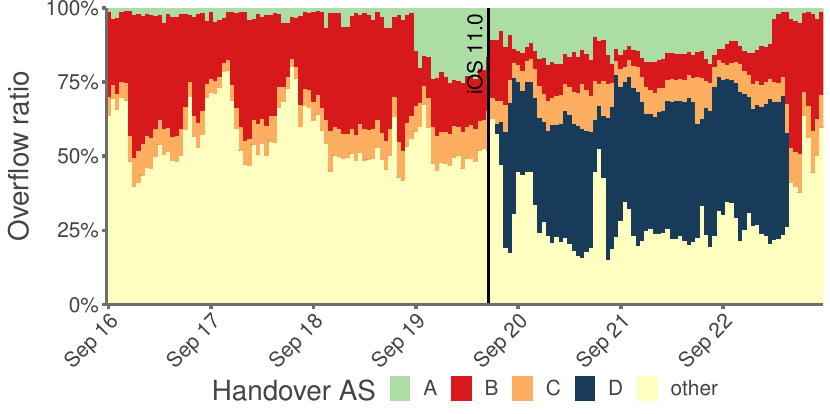}
  \caption{Overflow by handover AS during the iOS update.}
  \label{fig:apple-overflow-surprise}
\end{figure}

Finally, we turn our attention to {\em overflow}. 
This phenomenon is particularly hard for ISPs to handle as it is extremely difficult to predict this traffic.
We extend the overflow analysis by enhancing the traffic flows with handover AS information by leveraging SNMP for handover AS classification and determing the capacity of the of the peering links.
Figure~\ref{fig:apple-overflow-surprise} depicts the traffic delivered by Limelight via indirect links (i.e., via Other ASes as in Figure~\ref{fig:offloading-overflow}).
Limelight carries a significant amount of update traffic and is, at the same time, an offload \ac{cdn}.
Note that we grouped $\sim 40$ smaller \handoverAS{} into ``other'' as they do not change their behavior significantly.

We found the traffic to be stable regarding its \handoverAS{} distribution in the days before and after the update. 
However, on Sep. 19, \asA spikes in overflow traffic.
We assume that this is the pre-cache fill to distribute the content.
From an ISPs perspective, this tempts the assumption that \asA might carry a bulk of the update traffic.
However, when the actual update is delivered, \asB---an \ac{as} not seen before---spikes to deliver more than 40\% of the \overflow{} traffic due to Limelight's sudden use of servers in or behind \asB.
This \ac{as} is connected to the ISP via four direct connections, two of which become entirely saturated at peak times.
After three days, Limelight decides to no longer use these caches, and the \textit{normal} traffic pattern returns.
From the perspective of \asB, this could mean a multifold increase of their monthly bill, because the prevalent 95/5 billing~\cite{on95Billing} is affected by the traffic spike induced by Limelight.

\takeaway{Due to \offload and \overflow, traffic predictions during high-stress situations are becoming volatile, which leads to unexpected traffic behaviors and overloads on unrelated peering links}

\vspace{1em}
\section{Conclusion}
This paper assessed self-operated Meta-CDN deployments, exemplified by the first characterization of the \applecdn{}.
We shed light on Apple's DNS-based request mapping infrastructure by tracing it from over 800 vantage points.
We found Apple to use two major CDNs, Akamai and Lighlight, as well as its own content delivery infrastructure. 
Our findings are two-fold: First, we detected 34 sites cache-sites, and second, we determined the internal structure of the cache sites by analyzing HTTP headers.

We performed detailed measurements of the Apple Meta-CDN's behavior before, during, and after a flash crowd event caused by a major iOS software update in Sep. 2017.
Our analysis shows Apple handles the flash crowd on a per continent basis.
We found that in Europe, Apple offloads traffic to third-party \acp{cdn}. 
For the analysis of our traces from a European Eyeball ISP we define two behaviors, \offload{} and \overflow{}. 
Through these, we show the difficulty of an ISP in predicting which CDN of the Meta-CDN will be delivering the content due to unknown offload strategies.
Furthermore, we show that overflow traffic has a significant impact on unrelated peering connections, fully saturating links not expected to be affected. 
Since this traffic peak is unpredictable for an ISP, we argue that studying Meta-CDNs and their consequences on ISP traffic pose an interesting perspective for future work.

\begin{acks}
This work has been funded by the German Research Foundation (DFG) as part of project B1 and C2 within the Collaborative Research Center (CRC) 1053 -- MAKI.
The authors would like to thank the RIPE Atlas team, Marc Werner, Matthias Hollick, and David Hausheer for their valuable support during the measurement campaign.
We further thank the reviewers and our shepherd Philipp Richter for their valuable input and feedback.
\end{acks}

\clearpage
\bibliographystyle{ACM-Reference-Format}

\begin{thebibliography}{30}


\ifx \showCODEN    \undefined \def \showCODEN     #1{\unskip}     \fi
\ifx \showDOI      \undefined \def \showDOI       #1{#1}\fi
\ifx \showISBNx    \undefined \def \showISBNx     #1{\unskip}     \fi
\ifx \showISBNxiii \undefined \def \showISBNxiii  #1{\unskip}     \fi
\ifx \showISSN     \undefined \def \showISSN      #1{\unskip}     \fi
\ifx \showLCCN     \undefined \def \showLCCN      #1{\unskip}     \fi
\ifx \shownote     \undefined \def \shownote      #1{#1}          \fi
\ifx \showarticletitle \undefined \def \showarticletitle #1{#1}   \fi
\ifx \showURL      \undefined \def \showURL       {\relax}        \fi
\providecommand\bibfield[2]{#2}
\providecommand\bibinfo[2]{#2}
\providecommand\natexlab[1]{#1}
\providecommand\showeprint[2][]{arXiv:#2}

\bibitem[\protect\citeauthoryear{??}{ced}{[n. d.]}]%
        {cedexis}
 \bibinfo{year}{[n. d.]}\natexlab{}.
\newblock \bibinfo{title}{{Cedexis}}.
\newblock \bibinfo{howpublished}{\url{https://www.cedexis.com/}}.
\newblock


\bibitem[\protect\citeauthoryear{??}{con}{[n. d.]}]%
        {conviva}
 \bibinfo{year}{[n. d.]}\natexlab{}.
\newblock \bibinfo{title}{{Conviva}}.
\newblock \bibinfo{howpublished}{\url{https://www.conviva.com/}}.
\newblock


\bibitem[\protect\citeauthoryear{??}{loc}{[n. d.]}]%
        {locode}
 \bibinfo{year}{[n. d.]}\natexlab{}.
\newblock \bibinfo{title}{{UN/LOCODE Location Codes}}.
\newblock
  \bibinfo{howpublished}{\url{http://www.unece.org/cefact/codesfortrade/codes_index.html}}.
\newblock


\bibitem[\protect\citeauthoryear{??}{apn}{[n. d.]}]%
        {apnicixp}
 \bibinfo{year}{[n. d.]}\natexlab{}.
\newblock \bibinfo{booktitle}{\emph{{Visible ASNs: Customer Populations
  (Est.)}}}.
\newblock \bibinfo{publisher}{APNIC Labs}.
\newblock
\urldef\tempurl%
\url{https://stats.labs.apnic.net/aspop/}
\showURL{%
\tempurl}
\newblock
\shownote{Accessed: 2018-9-27.}


\bibitem[\protect\citeauthoryear{Adhikari, Guo, Hao, Hilt, and Zhang}{Adhikari
  et~al\mbox{.}}{2012a}]%
        {adhikari2012tale}
\bibfield{author}{\bibinfo{person}{Vijay~Kumar Adhikari}, \bibinfo{person}{Yang
  Guo}, \bibinfo{person}{Fang Hao}, \bibinfo{person}{Volker Hilt}, {and}
  \bibinfo{person}{Zhi-Li Zhang}.} \bibinfo{year}{2012}\natexlab{a}.
\newblock \showarticletitle{{A tale of three CDNs: An active measurement study
  of Hulu and its CDNs}}. In \bibinfo{booktitle}{\emph{INFOCOM Workshops}}.
\newblock


\bibitem[\protect\citeauthoryear{Adhikari, Jain, Chen, and Zhang}{Adhikari
  et~al\mbox{.}}{2012b}]%
        {VivisectingYoutube}
\bibfield{author}{\bibinfo{person}{Vijay~Kumar Adhikari},
  \bibinfo{person}{Sourabh Jain}, \bibinfo{person}{Yingying Chen}, {and}
  \bibinfo{person}{Zhi-Li Zhang}.} \bibinfo{year}{2012}\natexlab{b}.
\newblock \showarticletitle{{Vivisecting YouTube: An active measurement
  study}}. In \bibinfo{booktitle}{\emph{IEEE INFOCOM}}.
\newblock


\bibitem[\protect\citeauthoryear{Apple}{Apple}{[n. d.]a}]%
        {apple_softwareupdateurl}
\bibfield{author}{\bibinfo{person}{Apple}.} \bibinfo{year}{[n.
  d.]}\natexlab{a}.
\newblock \bibinfo{title}{{iOS SoftwareUpdate File URL}}.
\newblock
  \bibinfo{howpublished}{\url{http://mesu.apple.com/assets/com_apple_MobileAsset_SoftwareUpdate/com_
  apple_MobileAsset_SoftwareUpdate.xml}}.
\newblock


\bibitem[\protect\citeauthoryear{Apple}{Apple}{[n. d.]b}]%
        {apple_updatebrainurl}
\bibfield{author}{\bibinfo{person}{Apple}.} \bibinfo{year}{[n.
  d.]}\natexlab{b}.
\newblock \bibinfo{title}{{iOS UpdateBrain File URL}}.
\newblock
  \bibinfo{howpublished}{\url{http://mesu.apple.com/assets/com_apple_MobileAsset_MobileSoftwareUpdate_
  UpdateBrain/com_apple_MobileAsset_MobileSoftwareUpdate_UpdateBrain.xml}}.
\newblock


\bibitem[\protect\citeauthoryear{Bendfeldt, Blendin, Poese, Koldehofe,
  Hohlfeld, and {Ripe Atlas}}{Bendfeldt et~al\mbox{.}}{2017}]%
        {RipeAtlasios11}
\bibfield{author}{\bibinfo{person}{Fabrice Bendfeldt},
  \bibinfo{person}{Jeremias Blendin}, \bibinfo{person}{Ingmar Poese},
  \bibinfo{person}{Boris Koldehofe}, \bibinfo{person}{Oliver Hohlfeld}, {and}
  \bibinfo{person}{{Ripe Atlas}}.} \bibinfo{year}{2017}\natexlab{}.
\newblock \bibinfo{title}{{RIPE Atlas Measurement \#9299652: Apple iOS 11
  Release Day DNS Resolution Measurements of appldnld.apple.com.}}
\newblock
  \bibinfo{howpublished}{\url{https://atlas.ripe.net/measurements/9299652}}.
\newblock


\bibitem[\protect\citeauthoryear{Calder, Flavel, Katz-Bassett, Mahajan, and
  Padhye}{Calder et~al\mbox{.}}{2015}]%
        {calder2015anycast}
\bibfield{author}{\bibinfo{person}{Matt Calder}, \bibinfo{person}{Ashley
  Flavel}, \bibinfo{person}{Ethan Katz-Bassett}, \bibinfo{person}{Ratul
  Mahajan}, {and} \bibinfo{person}{Jitendra Padhye}.}
  \bibinfo{year}{2015}\natexlab{}.
\newblock \showarticletitle{{Analyzing the Performance of an Anycast CDN}}. In
  \bibinfo{booktitle}{\emph{Proceedings of IMC}}.
\newblock


\bibitem[\protect\citeauthoryear{Chen, Sitaraman, and Torres}{Chen
  et~al\mbox{.}}{2015}]%
        {chen2015mapping}
\bibfield{author}{\bibinfo{person}{Fangfei Chen}, \bibinfo{person}{Ramesh~K.
  Sitaraman}, {and} \bibinfo{person}{Marcelo Torres}.}
  \bibinfo{year}{2015}\natexlab{}.
\newblock \showarticletitle{{End-User Mapping: Next Generation Request Routing
  for Content Delivery}}. In \bibinfo{booktitle}{\emph{ACM SIGCOMM}}.
\newblock


\bibitem[\protect\citeauthoryear{Dimitropoulos, Hurley, Kind, and
  Stoecklin}{Dimitropoulos et~al\mbox{.}}{2009}]%
        {on95Billing}
\bibfield{author}{\bibinfo{person}{Xenofontas Dimitropoulos},
  \bibinfo{person}{Paul Hurley}, \bibinfo{person}{Andreas Kind}, {and}
  \bibinfo{person}{Marc~Ph. Stoecklin}.} \bibinfo{year}{2009}\natexlab{}.
\newblock \showarticletitle{On the 95-Percentile Billing Method}. In
  \bibinfo{booktitle}{\emph{Proceedings of PAM}}.
\newblock


\bibitem[\protect\citeauthoryear{Dobrian, Sekar, Awan, Stoica, Joseph, Ganjam,
  Zhan, and Zhang}{Dobrian et~al\mbox{.}}{2011}]%
        {dobrian2011conviva}
\bibfield{author}{\bibinfo{person}{Florin Dobrian}, \bibinfo{person}{Vyas
  Sekar}, \bibinfo{person}{Asad Awan}, \bibinfo{person}{Ion Stoica},
  \bibinfo{person}{Dilip Joseph}, \bibinfo{person}{Aditya Ganjam},
  \bibinfo{person}{Jibin Zhan}, {and} \bibinfo{person}{Hui Zhang}.}
  \bibinfo{year}{2011}\natexlab{}.
\newblock \showarticletitle{{Understanding the Impact of Video Quality on User
  Engagement}}. In \bibinfo{booktitle}{\emph{Proceedings of SIGCOMM}}.
\newblock


\bibitem[\protect\citeauthoryear{Flavel, Mani, Maltz, Holt, Liu, Chen, and
  Surmachev}{Flavel et~al\mbox{.}}{2015}]%
        {flavel2015fastroute}
\bibfield{author}{\bibinfo{person}{Ashley Flavel},
  \bibinfo{person}{Pradeepkumar Mani}, \bibinfo{person}{David Maltz},
  \bibinfo{person}{Nick Holt}, \bibinfo{person}{Jie Liu},
  \bibinfo{person}{Yingying Chen}, {and} \bibinfo{person}{Oleg Surmachev}.}
  \bibinfo{year}{2015}\natexlab{}.
\newblock \showarticletitle{{FastRoute: A Scalable Load-Aware Anycast Routing
  Architecture for Modern CDNs}}. In \bibinfo{booktitle}{\emph{USENIX NSDI}}.
\newblock


\bibitem[\protect\citeauthoryear{Frank, Poese, Lin, Smaragdakis, Feldmann,
  Maggs, Rake, Uhlig, and Weber}{Frank et~al\mbox{.}}{2013}]%
        {frank2013pushing}
\bibfield{author}{\bibinfo{person}{Benjamin Frank}, \bibinfo{person}{Ingmar
  Poese}, \bibinfo{person}{Yin Lin}, \bibinfo{person}{Georgios Smaragdakis},
  \bibinfo{person}{Anja Feldmann}, \bibinfo{person}{Bruce Maggs},
  \bibinfo{person}{Jannis Rake}, \bibinfo{person}{Steve Uhlig}, {and}
  \bibinfo{person}{Rick Weber}.} \bibinfo{year}{2013}\natexlab{}.
\newblock \showarticletitle{{Pushing CDN-ISP collaboration to the limit}}.
\newblock \bibinfo{journal}{\emph{ACM SIGCOMM CCR}} \bibinfo{volume}{43},
  \bibinfo{number}{3} (\bibinfo{year}{2013}), \bibinfo{pages}{34--44}.
\newblock


\bibitem[\protect\citeauthoryear{Gerber and Doverspike}{Gerber and
  Doverspike}{2011}]%
        {gerber2011backbone}
\bibfield{author}{\bibinfo{person}{Alexandre Gerber} {and}
  \bibinfo{person}{Robert Doverspike}.} \bibinfo{year}{2011}\natexlab{}.
\newblock \showarticletitle{{Traffic Types and Growth in Backbone Networks}}.
  In \bibinfo{booktitle}{\emph{IEEE OFC/NFOEC}}.
\newblock


\bibitem[\protect\citeauthoryear{Hohlfeld, R\"uth, Wolsing, and
  Zimmermann}{Hohlfeld et~al\mbox{.}}{2018}]%
        {cedexis-pam}
\bibfield{author}{\bibinfo{person}{Oliver Hohlfeld}, \bibinfo{person}{Jan
  R\"uth}, \bibinfo{person}{Konrad Wolsing}, {and} \bibinfo{person}{Torsten
  Zimmermann}.} \bibinfo{year}{2018}\natexlab{}.
\newblock \showarticletitle{{Characterizing a Meta-CDN}}. In
  \bibinfo{booktitle}{\emph{Proceedings of PAM}}.
\newblock


\bibitem[\protect\citeauthoryear{Inc.}{Inc.}{2017}]%
        {AppleAnnual16}
\bibfield{author}{\bibinfo{person}{Apple Inc.}}
  \bibinfo{year}{2017}\natexlab{}.
\newblock \bibinfo{title}{{2016 Annual Report}}.
\newblock
  \bibinfo{howpublished}{\url{http://files.shareholder.com/downloads/AAPL/3135363200x0xS1628280-16-20309/320193/filing.pdf}}.
\newblock


\bibitem[\protect\citeauthoryear{Labovitz, Iekel-Johnson, McPherson, Oberheide,
  and Jahanian}{Labovitz et~al\mbox{.}}{2010}]%
        {labovitz2010internet}
\bibfield{author}{\bibinfo{person}{Craig Labovitz}, \bibinfo{person}{Scott
  Iekel-Johnson}, \bibinfo{person}{Danny McPherson}, \bibinfo{person}{Jon
  Oberheide}, {and} \bibinfo{person}{Farnam Jahanian}.}
  \bibinfo{year}{2010}\natexlab{}.
\newblock \showarticletitle{{Internet Inter-domain Traffic}}. In
  \bibinfo{booktitle}{\emph{Proceedings of SIGCOMM}}.
\newblock


\bibitem[\protect\citeauthoryear{Liu, Wang, Yang, Wang, and Tian}{Liu
  et~al\mbox{.}}{2012}]%
        {liu2012optimizing}
\bibfield{author}{\bibinfo{person}{Hongqiang~Harry Liu}, \bibinfo{person}{Ye
  Wang}, \bibinfo{person}{Yang~Richard Yang}, \bibinfo{person}{Hao Wang}, {and}
  \bibinfo{person}{Chen Tian}.} \bibinfo{year}{2012}\natexlab{}.
\newblock \showarticletitle{{Optimizing Cost and Performance for Content
  Multihoming}}. In \bibinfo{booktitle}{\emph{Proceedings of SIGCOMM}}.
\newblock


\bibitem[\protect\citeauthoryear{{Michael Henriksen}}{{Michael Henriksen}}{[n.
  d.]}]%
        {aquatone}
\bibfield{author}{\bibinfo{person}{{Michael Henriksen}}.} \bibinfo{year}{[n.
  d.]}\natexlab{}.
\newblock \bibinfo{title}{{AQUATONE: A Tool for Domain Flyovers}}.
\newblock
  \bibinfo{howpublished}{\url{https://github.com/michenriksen/aquatone}}.
\newblock


\bibitem[\protect\citeauthoryear{Mukerjee, Bozkurt, Maggs, Seshan, and
  Zhang}{Mukerjee et~al\mbox{.}}{2016}]%
        {mukerjee2016broker}
\bibfield{author}{\bibinfo{person}{Matthew~K. Mukerjee},
  \bibinfo{person}{Ilker~Nadi Bozkurt}, \bibinfo{person}{Bruce Maggs},
  \bibinfo{person}{Srinivasan Seshan}, {and} \bibinfo{person}{Hui Zhang}.}
  \bibinfo{year}{2016}\natexlab{}.
\newblock \showarticletitle{{The Impact of Brokers on the Future of Content
  Delivery}}. In \bibinfo{booktitle}{\emph{ACM HotNets}}.
\newblock


\bibitem[\protect\citeauthoryear{Mukerjee, Bozkurt, Ray, Maggs, Seshan, and
  Zhang}{Mukerjee et~al\mbox{.}}{2017}]%
        {BruceBroker17}
\bibfield{author}{\bibinfo{person}{Matthew~K. Mukerjee},
  \bibinfo{person}{Ilker~Nadi Bozkurt}, \bibinfo{person}{Devdeep Ray},
  \bibinfo{person}{Bruce~M. Maggs}, \bibinfo{person}{Srinivasan Seshan}, {and}
  \bibinfo{person}{Hui Zhang}.} \bibinfo{year}{2017}\natexlab{}.
\newblock \showarticletitle{{Redesigning CDN-Broker Interactions for Improved
  Content Delivery}}. In \bibinfo{booktitle}{\emph{ACM CoNEXT}}.
\newblock


\bibitem[\protect\citeauthoryear{Nygren, Sitaraman, and Sun}{Nygren
  et~al\mbox{.}}{2010}]%
        {nygren2010akamai}
\bibfield{author}{\bibinfo{person}{Erik Nygren}, \bibinfo{person}{Ramesh~K.
  Sitaraman}, {and} \bibinfo{person}{Jennifer Sun}.}
  \bibinfo{year}{2010}\natexlab{}.
\newblock \showarticletitle{{The Akamai Network: A Platform for
  High-performance Internet Applications}}.
\newblock \bibinfo{journal}{\emph{SIGOPS OS Rev.}} \bibinfo{volume}{44},
  \bibinfo{number}{3} (\bibinfo{date}{Aug.} \bibinfo{year}{2010}),
  \bibinfo{pages}{2--19}.
\newblock
\showISSN{0163-5980}


\bibitem[\protect\citeauthoryear{Otto, S\'{a}nchez, Rula, and Bustamante}{Otto
  et~al\mbox{.}}{2012}]%
        {otto2012cdn}
\bibfield{author}{\bibinfo{person}{John~S. Otto}, \bibinfo{person}{Mario~A.
  S\'{a}nchez}, \bibinfo{person}{John~P. Rula}, {and}
  \bibinfo{person}{Fabi\'{a}n~E. Bustamante}.} \bibinfo{year}{2012}\natexlab{}.
\newblock \showarticletitle{{Content Delivery and the Natural Evolution of DNS:
  Remote Dns Trends, Performance Issues and Alternative Solutions}}. In
  \bibinfo{booktitle}{\emph{Proceedings of IMC}}.
\newblock


\bibitem[\protect\citeauthoryear{Poese, Frank, Ager, Smaragdakis, and
  Feldmann}{Poese et~al\mbox{.}}{2010}]%
        {poese2010padis}
\bibfield{author}{\bibinfo{person}{Ingmar Poese}, \bibinfo{person}{Benjamin
  Frank}, \bibinfo{person}{Bernhard Ager}, \bibinfo{person}{Georgios
  Smaragdakis}, {and} \bibinfo{person}{Anja Feldmann}.}
  \bibinfo{year}{2010}\natexlab{}.
\newblock \showarticletitle{{Improving Content Delivery Using Provider-aided
  Distance Information}}. In \bibinfo{booktitle}{\emph{Proceedings of IMC}}.
\newblock


\bibitem[\protect\citeauthoryear{{Ripe Network Coordination Centre}}{{Ripe
  Network Coordination Centre}}{[n. d.]}]%
        {RipeAtlas}
\bibfield{author}{\bibinfo{person}{{Ripe Network Coordination Centre}}.}
  \bibinfo{year}{[n. d.]}\natexlab{}.
\newblock \bibinfo{title}{{RIPE Atlas Internet Data Collection System}}.
\newblock \bibinfo{howpublished}{\url{https://atlas.ripe.net/}}.
\newblock


\bibitem[\protect\citeauthoryear{Su, Choffnes, Kuzmanovic, and Bustamante}{Su
  et~al\mbox{.}}{2009}]%
        {su2009drafting}
\bibfield{author}{\bibinfo{person}{Ao-Jan Su}, \bibinfo{person}{David~R.
  Choffnes}, \bibinfo{person}{Aleksandar Kuzmanovic}, {and}
  \bibinfo{person}{Fabi\'{a}n~E. Bustamante}.} \bibinfo{year}{2009}\natexlab{}.
\newblock \showarticletitle{{Drafting Behind Akamai: Inferring Network
  Conditions Based on CDN Redirections}}.
\newblock \bibinfo{journal}{\emph{IEEE/ACM ToN}} \bibinfo{volume}{17},
  \bibinfo{number}{6} (\bibinfo{year}{2009}), \bibinfo{pages}{1752--1765}.
\newblock


\bibitem[\protect\citeauthoryear{Wichtlhuber, Kessler, Bücker, Poese, Blendin,
  Koch, and Hausheer}{Wichtlhuber et~al\mbox{.}}{2017}]%
        {wichtlhuber17}
\bibfield{author}{\bibinfo{person}{Matthias Wichtlhuber}, \bibinfo{person}{Jan
  Kessler}, \bibinfo{person}{Sebastian Bücker}, \bibinfo{person}{Ingmar
  Poese}, \bibinfo{person}{Jeremias Blendin}, \bibinfo{person}{Christian Koch},
  {and} \bibinfo{person}{David Hausheer}.} \bibinfo{year}{2017}\natexlab{}.
\newblock \showarticletitle{{SoDA: Enabling CDN-ISP Collaboration with Software
  Defined Anycast}}. In \bibinfo{booktitle}{\emph{IFIP Networking}}.
\newblock


\bibitem[\protect\citeauthoryear{Xue, Choffnes, and Wang}{Xue
  et~al\mbox{.}}{2017}]%
        {xue2017cdn}
\bibfield{author}{\bibinfo{person}{Jing’an Xue}, \bibinfo{person}{David
  Choffnes}, {and} \bibinfo{person}{Jilong Wang}.}
  \bibinfo{year}{2017}\natexlab{}.
\newblock \showarticletitle{{CDNs Meet CN An Empirical Study of CDN Deployments
  in China}}.
\newblock \bibinfo{journal}{\emph{IEEE Access}}  \bibinfo{volume}{5}
  (\bibinfo{year}{2017}), \bibinfo{pages}{5292--5305}.
\newblock


\end{thebibliography}
\balance

\end{document}